\documentclass[twocolumn,prl,aps,superscriptaddress]{revtex4-1}
\usepackage[latin9]{inputenc}
\usepackage{color}
\usepackage{amstext}
\usepackage{amssymb}
\usepackage{graphicx}
\usepackage[unicode=true,pdfusetitle,
 bookmarks=false,
 breaklinks=false,pdfborder={0 0 1},backref=false,colorlinks=false]
 {hyperref}
\hypersetup{
 bookmarksnumbered=false,bookmarksopen=false}

\makeatletter
\@ifundefined{textcolor}{}
{%
 \definecolor{BLACK}{gray}{0}
 \definecolor{WHITE}{gray}{1}
 \definecolor{RED}{rgb}{1,0,0}
 \definecolor{GREEN}{rgb}{0,1,0}
 \definecolor{BLUE}{rgb}{0,0,1}
 \definecolor{CYAN}{cmyk}{1,0,0,0}
 \definecolor{MAGENTA}{cmyk}{0,1,0,0}
 \definecolor{YELLOW}{cmyk}{0,0,1,0}
}

\usepackage{amsmath}
\usepackage{graphicx}
\usepackage{amssymb}
\usepackage{txfonts}

\makeatother

\begin{document}
\title{Shaped Pulses for Energy Efficient High-Field NMR at the Nanoscale}

\author{J. Casanova}
\affiliation{Institut f\"ur Theoretische Physik and IQST, Albert-Einstein-Allee 11, Universit\"at Ulm, D-89069 Ulm, Germany}
\author{Z.-Y. Wang}
\affiliation{Institut f\"ur Theoretische Physik and IQST, Albert-Einstein-Allee 11, Universit\"at Ulm, D-89069 Ulm, Germany}
\author{I. Schwartz}
\affiliation{Institut f\"ur Theoretische Physik and IQST, Albert-Einstein-Allee 11, Universit\"at Ulm, D-89069 Ulm, Germany}
\author{M. B. Plenio}
\affiliation{Institut f\"ur Theoretische Physik and IQST, Albert-Einstein-Allee 11, Universit\"at Ulm, D-89069 Ulm, Germany}

\begin{abstract}
The realisation of optically detected magnetic resonance via nitrogen vacancy centers in diamond faces challenges
at high magnetic fields which include growing energy consumption of control pulses as well as decreasing sensitivities. Here
we address these challenges with the design of shaped pulses in microwave control sequences that achieve orders magnitude reductions in energy consumption and concomitant increases in sensitivity when compared to standard top-hat microwave
pulses. The method proposed here is general and can be applied to any quantum sensor subjected to pulsed control sequences.
\end{abstract}

\maketitle

\section{Introduction} Nuclear magnetic resonance (NMR) techniques~\cite{Rabi38} applied
to macroscopic samples have enabled fundamental scientific breakthroughs in organic chemistry, biology, medicine and material science \cite{Ernst87, Findeisen14}. Recently, NMR detection has been extended to the nanoscale ~\cite{SchmittGS17,BossCZ17,GlennBL18} where macroscopic detecting
coils \cite{Levitt08} are replaced by a quantum sensor based on the nitrogen-vacancy (NV) center in
diamond \cite{Doherty13, Dobrovitski13,Wu16}. These minute detectors can be located very close to
the sample under study thus opening the door for the detection of NMR signals emitted by nanoscale samples
or even by individual  nuclei \cite{Kolkowitz12, Taminiau12, Zhao12, Muller14, Lovchinsky16, Abobeih18}.
NV centers are particularly well suited for this purpose because their magnetic sub-levels can be
initialised and read-out with laser fields, while their hyperfine spin transitions are manipulated
entirely with microwave (MW) radiation~\cite{Doherty13, Dobrovitski13}. In addition, even at room 
temperature NV centers can achieve long coherent times thanks to the application of dynamical decoupling 
(DD) techniques~\cite{Ryan10,NaydenovDH11,Souza12}. Here, external microwave driving fields act continuously, 
or stroboscopically in the form of $\pi$-pulses, on the NV quantum sensor to average out environmental 
noise while preserving the sensitivity for external signals.

Typically sensing experiments based on NV centers are performed at relatively low static magnetic
fields, on the order (or lower) than a few hundred of Gauss~\cite{Gurudev07, Robledo11, Sar12, Zhao12,
Souza12, Kolkowitz12, Taminiau12, Liu13, Taminiau14, Waldherr14, Cramer16, Hensen15, Abobeih18},
with singular exceptions as, for example,~\cite{Muller14} and~\cite{Neumann10} that operate at
thousands of Gauss. For  nanoscale NMR the realisation of detection
under large magnetic fields (on the order of several Tesla) presents a number of 
advantages. These include longer nuclear $T_1$-times~\cite{Reynhardt01}, increased thermal
spin polarisation which leads to enhanced NMR signal strength, as well as larger chemical shifts which are key quantities in molecular structure determination \cite{Levitt08}. However, large magnetic fields also poses significant challenges caused by the increase of the nuclear Larmor frequency of the target nuclei. For continuous microwave driving, one requires the application on the NV of a microwave Rabi frequency equal to the nuclear Larmor-frequency to achieve the Hartmann-Hahn resonance condition~\cite{Hartmann62, Cai12}. Pulsed schemes give access to higher harmonics of the basic modulation frequency but at the cost of  reducing the effective NV-target coupling strength~\cite{Taminiau12}. Furthermore,  pulsed schemes assume the application of $\pi$-pulses on the NV in time intervals that are shorter than the nuclear Larmor frequency~\cite{Pasini08, Pasini08bis, Wang11}. A failure of this condition leads to severe reduction of the sensitivity to the NMR signal which, as we will show, scales as the inverse square of the Larmor frequency for 
fixed pulse duration. To restore the NMR signal, high peak power and high average power  should be delivered. Unfortunately high microwave
power lead to heating effects especially in biological samples \cite{Cao17}, and is difficult to achieve as microwave structures that deliver the control fields are limited in peak and average  power.

In this article we will first demonstrate these relationships between standard (top-hat like)
$\pi$-pulses of fixed length, the power requirements and the effective coupling strength
to the signal emanating from the target. Secondly, we present a solution to this problem based
on suitable shaping of long $\pi$-pulses which achieve an effective dynamics that has the same
effect as an instantaneous $\pi$-pulse restoring the ideal sensor-target interaction. In this 
manner we can extend the duration of the $\pi$-pulses
and reduce the required peak and average power to levels that are more
accessible to current technology and compatible with sensing applications in
biological samples~\cite{Wu16,Cao17}. Our protocol is universal and can be incorporated into any pulse sequence used in experiments to reduce microwave power consumption. Furthermore, our method is not restricted to NV centers and extends to a broad range of systems that benefit from DD sequences to reduce their noise level, e.g. trapped ions and a variety of solid state physics architectures, thus opening the field of DD under the critical limitation of accessible power.
 
\section{Preliminaries} We consider the detection of nuclear spins  at
a strong magnetic field $B_z\gtrsim 1$~T. If the Rabi frequency of the MW driving field
is  limited, then, for sufficiently high $B_z$,  nuclear spins complete several oscillations during a $\pi$-pulse. Now we 
analyse the reduction in sensitivity due to this effect. The Hamiltonian of an NV-nucleus system is
\begin{equation}\label{rfnucleus}
    H = DS_z^2 - \gamma_e B_z S_z -\omega_L I_z + S_z \ \vec{A}\cdot\vec{I}
        + \sqrt{2} \Omega S_x \cos(\omega t -\phi).
\end{equation}
Here, $S_z = |1\rangle\langle 1| - |-1\rangle\langle -1|$, $S_x =1/\sqrt{2}(|1\rangle\langle 0| 
+ |-1\rangle\langle 0| + {\rm H.c.})$, $D = (2\pi) \times 2.87$ GHz, 
$\gamma_e \approx -(2\pi) \times  28.024$ GHz/T and $\vec{A}$ is the hyperfine vector of 
the NV-nucleus interaction~\cite{Maze08}. For $B_z\gtrsim 1$~T, the NV energy splitting between 
the $|0\rangle \leftrightarrow |\pm 1\rangle$ transition is tens of GHz~\cite{Doherty13}
while $\omega_{L}$ (i.e. the nuclear Larmor frequency) would reach tens of MHz. The MW driving leads to 
$\sqrt{2} \Omega S_x \cos(\omega t -\phi)$, $\sqrt{2}$ is introduced for
convenience. When two NV levels, e.g. $|0\rangle$ and $|1\rangle$, are selected as the NV qubit basis, and setting the driving field on
resonance with that $|0\rangle \leftrightarrow |1\rangle$ transition (i.e. $\omega = D + |\gamma_e|
B_z$) one finds that Eq.~(\ref{rfnucleus}) is (see Appendix~\ref{appfirst})
\begin{equation}\label{basic}
    H = \frac{F_z(t)}{2} \ \sigma_z \big[ A^\perp_x I_x \cos{(\omega_n t)} + A^\perp_y I_y \sin{(\omega_n t)} \big].
 \end{equation}
Here $F_z(t)$ is the modulation function that appears as a consequence  of the MW
pulse sequence, $\omega_n$ is the nuclear resonance frequency, and $A^{\perp}_{x,y}$  are electron-nucleus 
coupling constants (see Appendix~\ref{appfirst}). We consider periodic pulse sequences of period $T$ such that 
$F_z(t) = \sum_l f_l \cos{(l \omega_{\rm m} t)}$ where $\omega_{\rm m}= \frac{2\pi}{T}$ and $
f_l = \frac{2}{T}\int_0^T F_z(s)\cos{(l \omega_{\rm m} s)} \ ds$. Examples of these sequences are those of the XY family~\cite{Maudsley86, Gullion90} or more sophisticated schemes~\cite{Ryan10, Souza11, Casanova15, Wang16, Casanova16, Wang17, Casanova17, Haase17}. 
For $k\omega_{\rm m} = \omega_n$ (resonance condition with the $k$th harmonic, i.e. for $l=k$ ) Eq.~(\ref{basic}) is
\begin{equation}\label{nvnucleus}
H=\frac{f_k}{4} A^{\perp}_x \sigma_z I_x.
\end{equation}
The latter can be easily seen as Eq.~(\ref{basic}) can be expanded as 
\begin{eqnarray}
H&=&\frac{f_k}{2}\cos{(k \omega_{\rm m} t)}  \sigma_z \big[ A^\perp_x I_x \cos{(\omega_n t)} + A^\perp_y I_y \sin{(\omega_n t)} \big]\nonumber \\
&+&\sum_{l\neq k} \frac{f_l}{2}\cos{(l \omega_{\rm m} t)}  \sigma_z \big[ A^\perp_x I_x \cos{(\omega_n t)} + A^\perp_y I_y \sin{(\omega_n t)} \big].\nonumber \\
\end{eqnarray}
Now, if one uses the resonance condition $k\omega_{\rm m} = \omega_n$, the above Hamiltonian is approximated by 
\begin{eqnarray}\label{newbig}
H&\approx &\frac{f_k}{4} A^{\perp}_x \sigma_z I_x\nonumber \\
&+&\sum_{l\neq k} \frac{f_l}{8} \sigma_z  A^\perp_x I_x\big[e^{i(\omega_{n} \frac{k-l}{k} t)} + e^{-i(\omega_{n} \frac{k-l}{k} t)}\big],\nonumber \\
&+&\sum_{l\neq k} \frac{f_l}{8i} \sigma_z  A^\perp_y I_y\big[e^{i(\omega_{n} \frac{k-l}{k} t)} - e^{-i(\omega_{n} \frac{k-l}{k} t)}\big],
\end{eqnarray}
where we have eliminated counter-rotating terms. Now, in order to get Eq.~(\ref{basic}), the last two lines in Eq.~(\ref{newbig}) can be eliminated 
under the condition $|\gamma_n B_z| >> k |A^{\perp}_x|$. The latter can be derived by comparing the time-dependent phases and the couplings in the last two lines of Eq.~(\ref{newbig}), and noting that, for large $B_z$ fields, we have $\omega_{n} \approx \gamma_n B_z$. 
Note also that, the condition $|\gamma_n B_z| >> k |A^{\perp}_x|$ can be easily fulfilled in situations with a large $B_z$ field (which  
corresponds to our operating regime) as we will demonstrate in our numerical simulations.

Hence, according to Eq.~(\ref{nvnucleus}) it is the product of $A^{\perp}_x$ and the Fourier $f_k$ coefficient which determines 
the strength of the NV-nucleus interaction. If the target is a classical signal, e.g. an oscillating magnetic field of the kind $\vec{B}_s
\cos{(\omega_{\rm s} t)}$, the sensor target Hamiltonian in case of resonance with the $k$th 
harmonic (in this case $k \omega_{\rm m} = \omega_s$) is  
\begin{equation}\label{forclassical}
H = \frac{\Omega_s}{4}f_k \ \sigma_z
\end{equation}
with $\Omega_s$ the Rabi frequency associated to  the classical field (see Appendix \ref{appsecond}).

\begin{figure}[t]
\hspace{-0.45 cm}\includegraphics[width=1.05\columnwidth]{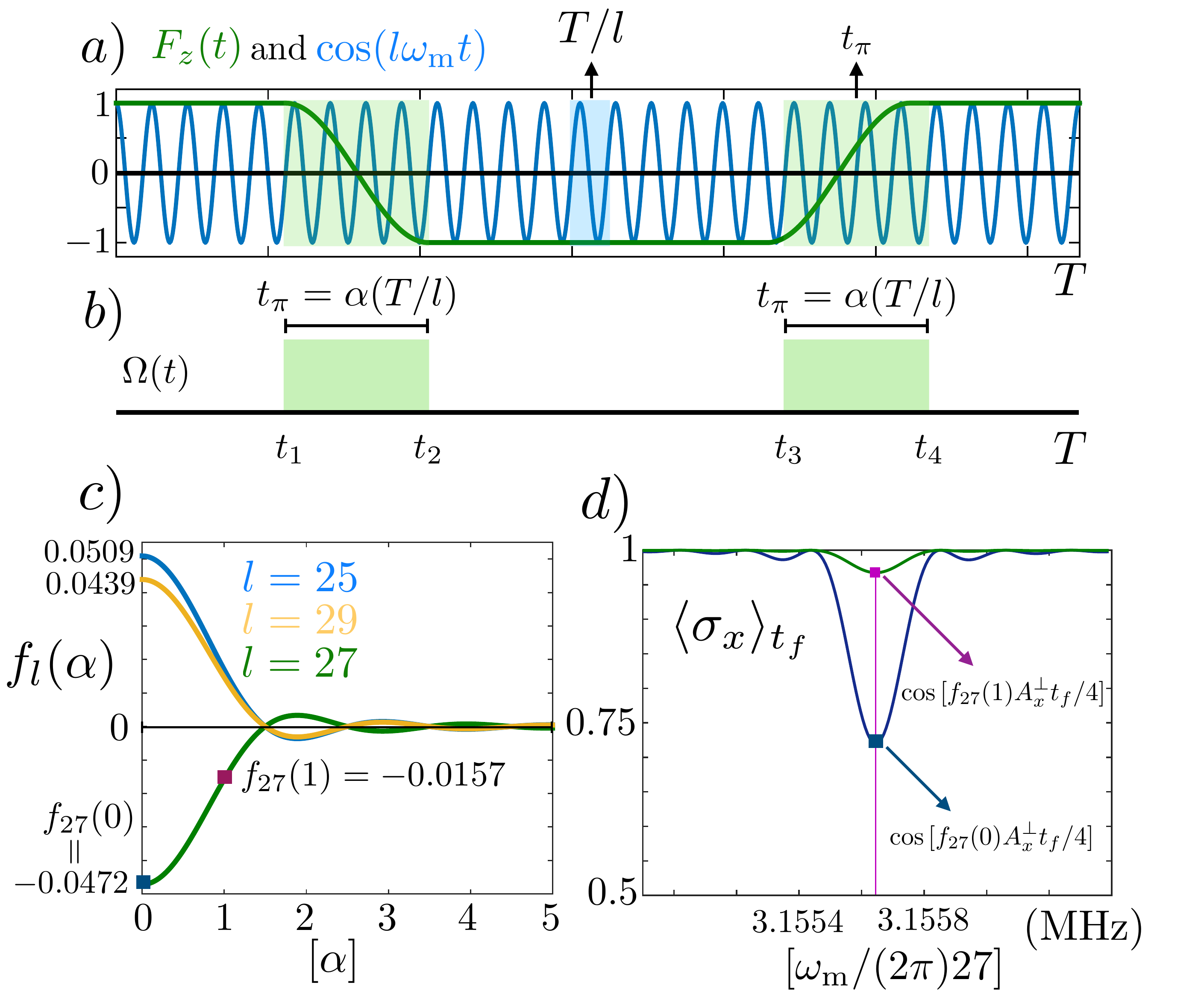}
\caption{ a) Modulation function $F_z(t)$ (green line) and its $l$th Fourier component $\cos(l \omega_{\rm m} t)$ (in blue)  for $l=27$. Superimposed (shaded in green) the MW pulses width. Shaded in blue it is shown one period, $T/l$, of  $\cos(l \omega_{\rm m} t)$, $T$ being the period of $F_z(t)$.  b) MW pulses (in green) where it can be seen the relation between the pulse width $t_\pi$, and the period of $\cos(l \omega_{\rm m} t)$. In this case we have $\alpha=4$. c) Decay of the $f_k(\alpha)$ coefficients with $\alpha$  for $l=25$ (blue), $27$ (green), and $29$ (yellow). We highlight the value of $f_{27}(1)$ that is used in the next plot. d) Expectation value of the NV $\sigma_x$ operator evolving under Hamiltonian~(\ref{basic}) with $B_z=2$ T and   $\vec{A}= (2\pi)\times [19.12, 55.21, -96.82]$ kHz. We applied 1120 $\pi$-pulses (final evolution time  $ t_f \approx 177 \ \mu$s). The blue line is the signal for ideal instantaneous pulses with large contrast corresponding to $f_{27}(0) = -\frac{4}{27 \pi  }\approx  -0.0472$, i.e. $\alpha=0.$ The green line corresponds to $\alpha=1$ (we simulated top-hat $\pi$-pulses with $\Omega \approx (2\pi)\times42$ MHz, leading to $f_{27}(1) = -0.0157$) which yields to a reduced contrast. Violet panel is the theoretical prediction of $\langle \sigma_x\rangle_{t_f} = \cos{[f_k(\alpha) A^{\perp}_x t_f/4]}$ for $f_{27}(1)$ (i.e. $k=27$ and $\alpha =1$), blue square is the prediction for instantaneous $\pi$-pulses ($\langle \sigma_x\rangle_{t_f} = \cos{[f_{27}(0) A^{\perp}_x t_f/4]}$).}
\label{tophat}
\end{figure}

\section{Signal reduction} For the common case of top-hat pulses, the value of each $f_l$ coefficient for the elementary block in Fig.~\ref{tophat} a) is
\begin{eqnarray}\label{integralbig}
    f_l &=&  \frac{2}{T}\big[\int_0^{t_1} \cos{(l \omega_{\rm m} s)} \ ds
        + \int_{t_1}^{t_2} \cos{[\Omega (s-t_1)]} \cos{(l \omega_{\rm m} s)} \ ds \nonumber \\
        & & - \int_{t_2}^{t_3} \cos{(l \omega_{\rm m} s)} \ ds - \int_{t_3}^{t_4} \cos{[\Omega (s-t_3)]} \cos{(l \omega_{\rm m} s)} \ ds \nonumber \\ && +\int_{t_4}^{T} \cos{(l \omega_{\rm m} s)} \ ds  \big],
\end{eqnarray}
where $t_2-t_1= t_4-t_3 \equiv t_{\pi}$ are the lengths of the $\pi$-pulses,
see~Fig.~\ref{tophat}~a). For instantaneous $\pi$-pulses, $t_{\pi}=0$, the
integrals $ \int_{t_1}^{t_2} \cos{[\Omega (s-t_1)]} \cos{(l \omega_{\rm m} s)} \ ds$
and $ \int_{t_3}^{t_4} \cos{[\Omega (s-t_1)]} \cos{(l \omega_{\rm m} s)} \ ds$
in Eq.~(\ref{integralbig}) vanish, and $|f_l| =|\frac{4}{\pi l}|$ for odd 
$l$, and $|f_l| = 0$ for even $l$. When $t_{\pi}$ is non-negligible
one finds
\begin{eqnarray}\label{analyticalsol}
 f_l\equiv f_l(\alpha) &=& \frac{4(-1)^{(l+1)/2}\cos{(\alpha\pi)}}{(4\alpha^2-1)l\pi},
\end{eqnarray}
which implies that the sensitivity under a finite-width pulse sequence decreases as $\alpha^{-2}$,
where $\alpha$ equals the length of $t_{\pi}=\alpha (T/l)$ measured in terms of 
the number of nuclear Larmor periods, Fig.~\ref{tophat} b). If we aim for 
a resonance at a certain $l$th harmonic we need to set $T = 2\pi l/\omega_n$ (equivalent 
to the resonance condition $l\omega_{\rm m} = \omega_{\rm n}$) where $\omega_n$, grows linearly with the applied magnetic field $B_z$. 
Regarding the latter, note that according to the expressions in Appendix~\ref{appfirst} we have  $\omega_n = |\vec{\omega}_n|$ where $\vec{\omega}_n = (-\frac{1}
{2} A_x, -\frac{1}{2} A_y, \omega_{\rm L} -\frac{1}{2} A_z)$ and $\omega_{\rm L} = \gamma_n B_z$. Then, we have $\omega_n \approx  
\omega_{\rm L} -\frac{1}{2} A_z$, and for the case of large $B_z$ fields the behaviour of the resonance frequency  $\omega_n$ can be well approximated by $\gamma_n B_z$.

From Eq.~(\ref{rfnucleus}) we have $t_{\pi} = \pi/{\Omega}$, hence we have
\begin{equation}\label{relation}
    \Omega = \frac{\pi}{\alpha} \frac{l}{T} \approx \frac{\gamma_n B_z}{2\alpha},
\end{equation}
where we have used that $T= 2\pi l/\omega_n$ and that, for large $B_z$ fields,  $\omega_n \approx \gamma_n B_z$.
Equation~(\ref{relation}) implies for  $B= 2$ T ($5$ T), a proton nuclear spin as a target ($\gamma_n\equiv\gamma_{\rm H} =
(2\pi)\times 42.57$ MHz/T),   and a peak power limited by a maximum achievable value for $\Omega$, namely $\Omega = (2\pi)\times10$ MHz, that $\alpha \approx 4.26 \ (10.65)$. Note that, the peak power is $\propto \Omega^2$ while average power is $\propto \Omega$, in this respect see Supplemental Material~\cite{Supplemental}.

In Figure~\ref{tophat} c) we show the rapid decay of $f_l(\alpha)$ coefficients with $\alpha$,
for $l=25$ (blue), $27$ (green), and $29$ (yellow) dictated by Eq.~(\ref{analyticalsol}). In Fig.~\ref{tophat} d)
we have computed the spectrum of a problem involving an NV center and a H nucleus for two values
of $\Omega$ with  $B_z = 2$ T  (see caption for details). In addition,
with Hamiltonian~(\ref{nvnucleus}) one can predict that, under the resonance condition, and assuming $\rho_0 = 
\frac{1}{4} (\mathbb{I} + \sigma_x)\otimes \mathbb{I}$ as the initial state of the NV-nucleus system  (i.e. the NV is initialised in an equal
superposition of the $|0\rangle$ and $|1\rangle$ states, and the nucleus is in a thermal state) the measured signal, i.e. the NV coherence, is  
\begin{equation}\label{nucsig}
\langle \sigma_x\rangle_{t_f} = \frac{1}{4}{\rm Tr}\bigg[ \rho_0 \ e^{i f_k(\alpha) A^{\perp}_x t_f/4 \sigma_z \sigma_x} \ (\sigma_x\otimes \mathbb{I}) \bigg] =  \cos  [f_k(\alpha) A^{\perp}_x t_f/4].
\end{equation}
with $t_f$ the final time of the sequence. In Fig~\ref{tophat} d), the vertical violet line corresponds to the theoretically predicted resonance 
according Eq.~(\ref{nucsig})  and we can observe how it matches with the numerically computed signal depth (green curve) for a finite value of
the MW Rabi frequency at the resonance position $\omega_{\rm m}= \omega_n/27$. The blue square in the same figure is the signal depth for 
instantaneous pulses (i.e. at infinite MW power) corresponding to use $f_{27}(0)$ in Eq.~(\ref{nucsig}). It is noteworthy to mention that, to take into account 
different effects as finite-width pulses as well as the presence of different rotating axes in the applied $\pi$-pulses (see next section), the 
numerical  simulations in this article have been performed starting from Hamiltonian~(\ref{forsimulations}) in Appendix~\ref{appfirst}. This Hamiltonian    
only assumes the elimination of fast counter rotating terms, of the order of GHz, on the MW driving as well as the presence of the third NV spin level which is detuned by a similar frequency amount of the order of GHz. For more details, see 
Eq.~(\ref{forsimulations}) and the paragraph below in Appendix~\ref{appfirst}. 

Hence, we can conclude that if an experiment is conducted at high $B_z$ (which implies large $\alpha$ to compensate for  limited $\Omega$)
the obtained signal gets dramatically reduced due to the decay of each $f_l(\alpha)$ with the $\alpha$ parameter.
\begin{figure}[t!]
\hspace{-0.2 cm}\includegraphics[width=1.0\columnwidth]{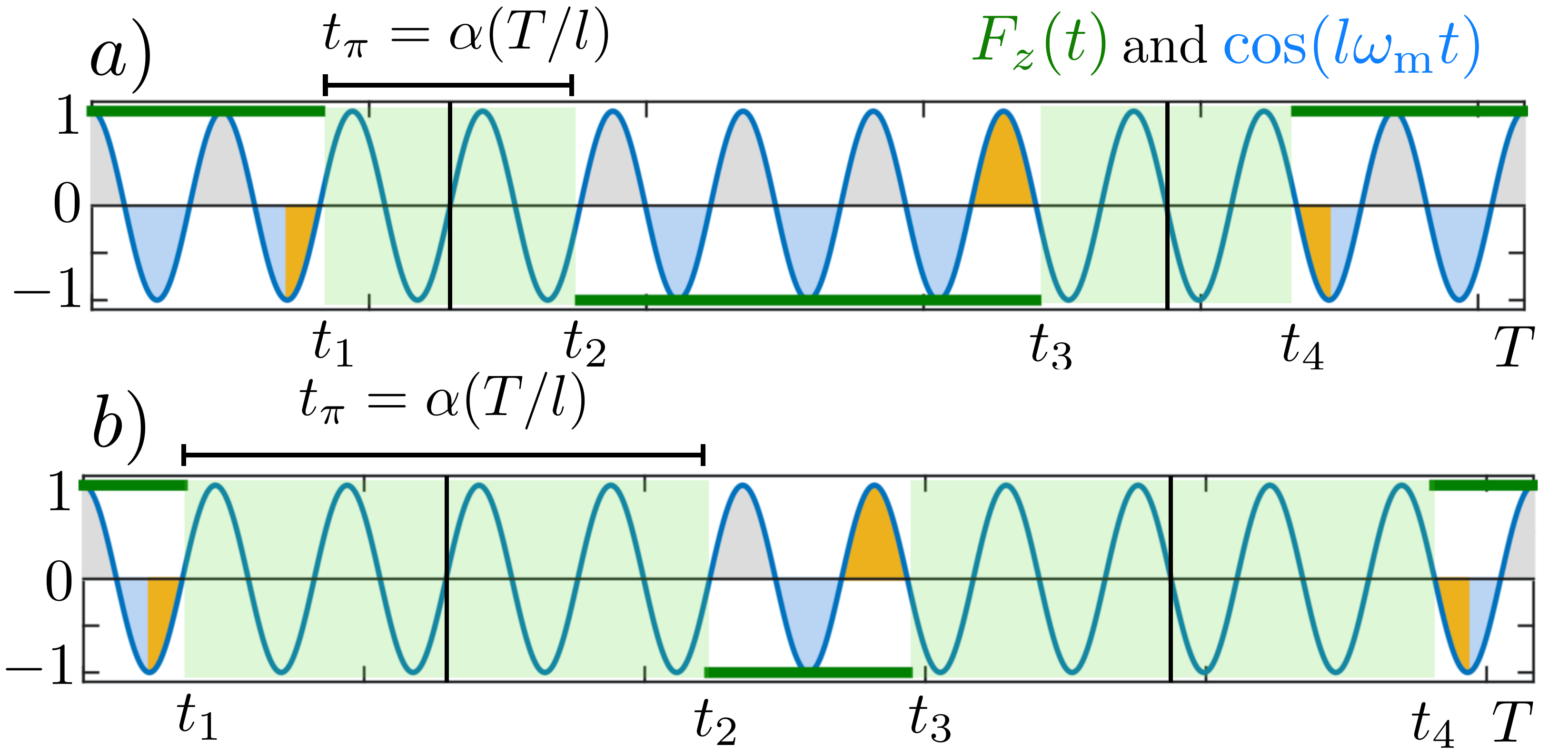}
\caption{$F_z(t)$ and  $\cos{(l\omega_{\rm m} t)}$  for two different pulse widths (shaded in green). a) $t_{\pi} = 2(T/l)$ and b)  $t_{\pi} = 4(T/l)$, in both cases $l=11$. In the $\pi$-pulse regions it is remarked the center of each finite-width $\pi$-pulse with a vertical black line. The latter would correspond with the locations of the instantaneous $\pi$-pulses.  The areas in yellow contribute to the value of $f_l$, while grey areas (positive oscillations) are cancelled by blue areas (negative oscillations).}
\label{equivalence}
\end{figure}
\section{Shaped $\pi$-pulses} If we measure the signal at a certain $l$th harmonic of $F_z(t)$ the length of the $\pi$-pulses  may extend
over many Larmor periods  as long as the extended $\pi$-pulse action equals
that of an instantaneous pulse. This can be seen in Fig.~\ref{equivalence}, where $F_z(t)$ (green flat lines), $\cos{(l\omega_{\rm m} t)}$
(solid blue lines) for $l=11$, and two possible time intervals used for the $\pi$-pulses
(shaded in green panels) are shown. Specifically in Fig.~\ref{equivalence} a)
$t_{\pi} = 2 (T/l)$, while in b) $t_{\pi} = 4 (T/l)$. Then, if one assumes that the
integral of $f_l$ could be written as 
\begin{eqnarray}\label{shapedfl}
f_l &=& \frac{2}{T}\int_0^T F(s)\cos{(l \omega_{\rm m} s)} \ ds\nonumber\\
&=&   \frac{2}{T}\bigg[\int_0^{t_1} \cos{(l \omega_{\rm m} s)} \ ds -\int_{t_2}^{t_3}
\cos{(l \omega_{\rm m} s)} \ ds  \nonumber\\
&+&\int_{t_4}^{T} \cos{(l \omega_{\rm m} s)} \ ds  \bigg],
\end{eqnarray}
i.e. without any contribution of the regions containing the $\pi$-pulses (see next paragraph 
for an explicit construction that will allow us having $f_l$ coefficients of the form that appears in Eq.~\ref{shapedfl}) both cases in Figs.~\ref{equivalence} a) and b) 
would lead to the same ideal value $|f_l| =|\frac{4}{\pi l}|$ as opposed to Eq. (\ref{analyticalsol}). This
is because when $\pi$-pulses contain a natural number of periods of $\cos{(l\omega_{\rm m} t)}$
the latter expression of $f_l$ is reduced to the integral of yellow areas in Figs.~\ref{equivalence}, which are equal for a) and b). This offers the opportunity to extend their length until $t_{\pi} = (l-1)/2(T/l)$ and  the potential to significantly reduce the Rabi frequency and hence the MW power.

Now, by substituting top-hat pulses for appropriately shaped pulses we can recover
the ideal $|f_l| =|\frac{4}{\pi l}|$ scaling.
This gains a factor of $\alpha^2$ in sensitivity and allows for a significant reduction in the power
requirements of the pulsed schemes. To this end we consider the shaped $\pi$-pulse Hamiltonian as
\begin{equation}\label{pulseH}
H_p = \frac{\Omega(t)}{2}(|1\rangle \langle 0| e^{i\phi} + |0\rangle \langle 1| e^{-i\phi}) = \frac{\Omega(t)}{2} \sigma_\phi,
\end{equation}
and a pulse width equal to a natural number of oscillations of $\cos{(l\omega_{\rm m} t)}$. 
\begin{figure}[b!]
\hspace{-0.4 cm}\includegraphics[width=1.04\columnwidth]{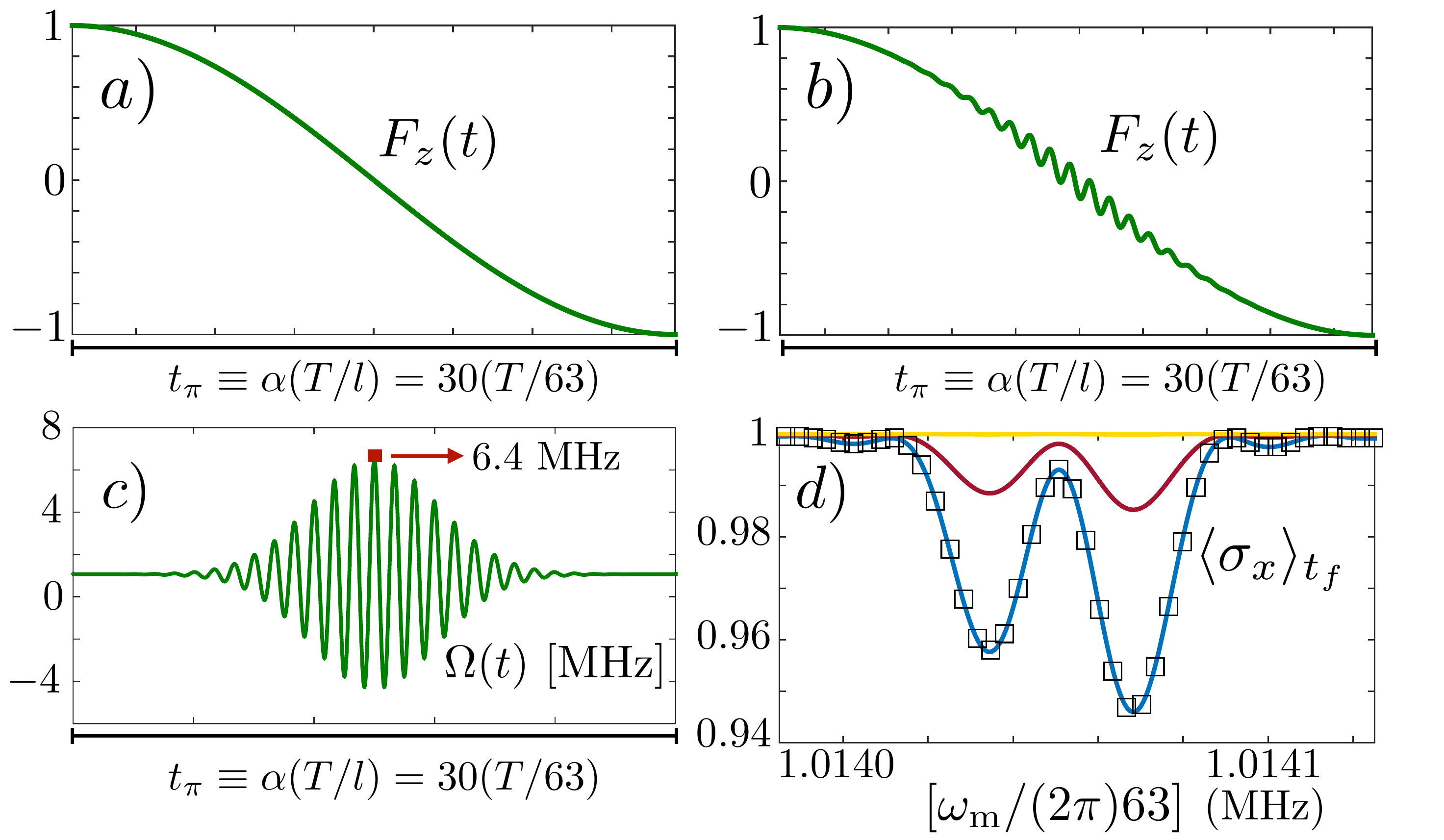}
\caption{a) $F_z(t)$ in the $\pi$-pulse region after the application of a constant $\Omega$, while in b) we can observe the oscillating character of $F_z(t)$ because of the action of  $\Omega(t)$ in c). The latter is displayed over $t_{\pi} = 30(T/63)\approx 0.469 \ \mu$s, i.e. $\alpha=30$ and $l=63$.  We selected $c=t_\pi/10$. The maximum value of $\Omega(t)$ is $\approx (2\pi)\times6.4$ MHz. d) Spectrum for ideal instantaneous pulses (blue), shaped pulses (squares), and top-hat pulses with a constant $\Omega$ of  $(2\pi)\times20$ MHz (yellow line) and $(2\pi)\times40$ MHz (red line).}
\label{modulatedRabi}
\end{figure}
In the rotating frame of $H_p$, a $\sigma_z$ electronic operator evolves, for the first shaped $\pi$-pulse,  as  $\sigma_z \rightarrow \cos{\bigg[\int_{t_1}^{t}  \Omega(s) \ ds\bigg]} \sigma_z + \sin{\bigg[\int_{t_1}^{t}  \Omega(s) \ ds\bigg]} \sigma_\phi^{\perp}$, where $t< t_1+t_{\pi}\equiv t_2$, and $\sigma_\phi^{\perp} = -i(|1\rangle \langle 0| e^{i\phi} - |0\rangle \langle 1| e^{-i\phi})$. This description is similar for any other shaped $\pi$-pulse of the sequence by simply replacing $t_1$ by $t_j$, the latter being the initial time of the $j$th shaped $\pi$-pulse. Now we focus only on the part containing the $\sigma_z$ operator, i.e. the one leading to the $F_z(t) \equiv \cos{\big[\int_{t_j}^{t_{j+1}} \Omega(s) \ ds\big]}$ modulation function, because the $\sigma_\phi^{\perp}$ component leads to the $F_x(t)$ and $F_y(t)$ modulation functions that do not contribute to the spectrum if the sequence contains pulses over different directions~\cite{Haase16,Lang17}. Then, we have to find a $\Omega(s)$ that minimise, in the shaped $\pi$-pulse region, the overlap between $F_z(t)$ and $\cos{(l \omega_{\rm m} t)}$, this is\vspace{-2 mm}
\begin{equation}\label{condition}
\int_{t_j}^{t_{j+1}} F_z(u) \ \cos{\big(l \omega_{\rm m} u\big)}  \ du = 0,
\end{equation}
with $t_{j+1} - t_{j} = t_{\pi}$. In addition, in order to have a continuous $F_z(t)$ the following boundary conditions are needed: $F_z(t_j) = (-1)^{n+1} \ \ {\rm and} \ \  F_z(t_{j+1}) = (-1)^n$, for the $n$th shaped $\pi$-pulse.

These conditions have an infinite number of solutions and here we present as an example
the analytical solution
\begin{equation}\label{ansatz}
    F_z(u) = \cos{\bigg[\frac{\pi}{t_\pi} (u - t_{j})\bigg]} -\beta e^{-\frac{(u-t_p)^2}{2c^2}} \sin{\bigg[\frac{2\pi \alpha }{t_\pi}(u - t_{j})\bigg]},
\end{equation}
with $\beta$ a parameter that will be fixed with Eq.~(\ref{condition}). For the first
shaped $\pi$-pulse $t_p$ is the middle point in between $t_1$ and $t_2$, or between $t_3$ and $t_4$
for the second one. The $\alpha$ and $c$ parameters can be advisedly adjusted, such
that their value determine the pulse length and maximum intensity of the employed $\Omega(t)$
in Eq.~(\ref{pulseH}). The above solution is valid for $\alpha$ equal to $1 ,2, 3, ..$ i.e.
when the shaped $\pi$-pulse contains a natural number of periods of $\cos{(l\omega_{\rm m} t)}$.
Equation~(\ref{condition}) leads to the following condition for $\beta$ (see Appendix~\ref{appthird})
\begin{equation}
\beta =  \frac{4\sqrt{2}\gamma \alpha}{(4\alpha^2-1)\pi^{3/2}}/\big[1 - \exp{\big(\frac{-8\alpha^2\pi^2}{\gamma^2}\big)}\big]
\end{equation}
where $\gamma$ and  $c$ are related
as $\gamma=t_\pi/c$. Once $\beta$ is chosen, one can calculate $\Omega(t)=\partial_t \arccos{[F(t)]}$
from  $F_z(t) = \cos{\big[\int_{t_j}^{t_{j+1}} \Omega(s) \ ds\big]}$.

Fig.~\ref{modulatedRabi} a) shows the evolution of $F_z(t)$ from $+1$ to $-1$ that results
from the application of a $\pi$-pulse  with a constant $\Omega$, i.e. a standard top-hat
pulse as those in Fig.~\ref{tophat} b), while Fig.~\ref{modulatedRabi} b) corresponds to $F_z(t)$
for the shaped $\pi$-pulse whose $\Omega(t)$ is shown in c). To further visualise the effect of the modulated Rabi frequency $\Omega(t)$ in the $F_z(t)$ function, in the Supplemental Material~\cite{Supplemental} we have included a whole period of $F_z(t)$ that results from the application of the $\Omega(t)$ in Fig.~\ref{modulatedRabi} c). In Fig.~\ref{modulatedRabi} d) we
have computed the response versus frequency that results from a system at $B_z=1.5$ T involving a NV center and two H nuclei with
$\vec{A}_1 = (2\pi) \times [0, 14.43, -46.63]$  kHz and,  $\vec{A}_2 = (2\pi) \times
[-10.93, 6.31, -42.34]$  kHz. The blue line corresponds to the signal obtained for $1440$
ideal instantaneous $\pi$-pulses (final time of the sequence $t_f\approx 0.71$ ms). The
black squares represent the signal that emerges when our shaped pulses in Fig.~\ref{modulatedRabi}
c)  are applied. Here we use the sequence [XYXYYXYX]$^{N}$ for its robustness against errors on the MW control (for an analysis including non-robust pulse constructions see Suplemental Material~\cite{Supplemental}) with $N=180$ and X (Y) a shaped pulse with  phase $\phi=0$ ($\phi=\frac{\pi}{2}$), see Eq.~(\ref{pulseH}). This signal overlaps with the ideal one  employing instantaneous $\pi$-pulses which demonstrates the efficiency of our method. At this point it is important  to clarify that the previous
numerical simulation involving NV nuclei  has been computed from Eq.~(\ref{forsimulations}) replacing the driving in that equation 
by the pulse Hamiltonian in Eq.~(\ref{pulseH}). 
The yellow line corresponds to the
application of standard top-hat $\pi$-pulses with $\Omega=(2\pi)\times20$ MHz, while the red
one uses $\Omega=(2\pi)\times40$ MHz. In these two cases, the signal contrast is seriously reduced with respect to the one obtained with our shaped
pulses  which employ a maximum of only $\Omega = (2\pi)\times 6.4 $ MHz.
\begin{figure}[t!]
\hspace{-0.2 cm}\includegraphics[width=1.0\columnwidth]{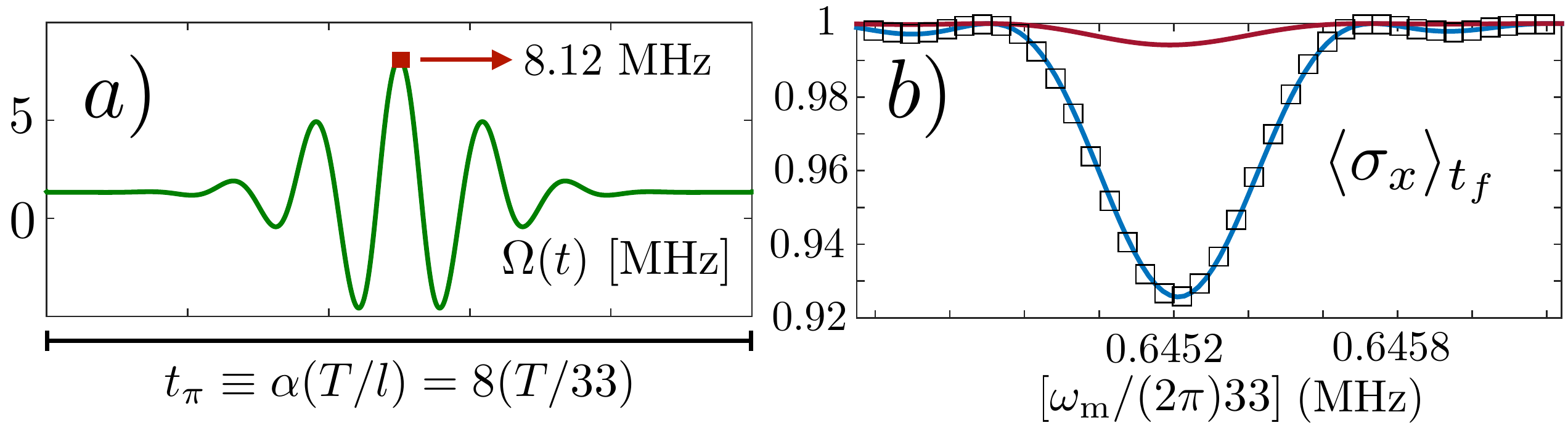}
\caption{Our method applied to a classical signal of the form $B_s [\cos{(\omega_{s_1}t)} + \cos{(\omega_{s_2}t)} ]$ with $\omega_{s_{1,2}} = (2\pi)\times (21.288, 21.295)$ MHz, and $B_s=0.2$ G. The modulated Rabi frequency in a)  leads to a signal in b) (squares) that overlaps with the ideal signal obtained with instantaneous  $\pi$-pulses (blue line). We used $\alpha=8$ and $c=t_\pi/10$. The red line in b) uses top-hat $\pi$-pulses with $\Omega = (2\pi)\times10$ MHz.}
\label{classicalsignal}
\end{figure}

In Fig.~\ref{classicalsignal} the same effect  is shown for a classical signal target. Here we demonstrate that the contrast reduction appears even at not so high $B_z$ fields (note we used a classical signal with frequencies corresponding to H nuclei at $B_z \approx 0.5$ T). In this respect, by inspecting Eq.~(\ref{relation}) it gets clear that finding a non negligible value for $\alpha$, which leads to a reduced  $f_l(\alpha)$, depends on the ratio between $B_z$ and $\Omega$. Hence, if the MW source cannot deliver high power, i.e. only low values of $\Omega$ are accesible, we get poor signal contrast. However, the  application of our shaped $\pi$-pulses  leads to an undistinguishable signal (squares) with respect to the ideal spectrum (blue line). The red line with poor contrast has been computed with top hat pulses corresponding to $\Omega = (2\pi)\times10$ MHz.
We used the same  $[{\rm XYXYYXYX}]^{N}$ sequence for $N=5$. That is 40 pulses, shaped and top-hat, have been applied.  The final time of the sequence is $t_f\approx0.03$ ms and shaped pulses are generated with a maximum of $\Omega = (2\pi)\times 8.12$ MHz.

Finally, we want to remark that in order to achieve similar results to  the ideal case in Fig.~\ref{modulatedRabi}
d) with finite top-hat pulses, a value for $\Omega$ of, at least, $\Omega = (2\pi)\times100$ MHz is needed. This is much larger than the maximum of $\Omega(t)$ in  Fig.~\ref{modulatedRabi} c) which implies that our method requires a much lower peak power. More specifically, the square ratio (peak power ratio) of these two quantities is $[(2\pi)\times100 / (2\pi)\times6.4]^2 \approx 244$. In addition, the ratio between the average energy required by the two approaches is $E_{[\pi-\rm pulse ]}^{\rm top-hat}/E_{[\pi-\rm pulse]}^{\rm shaped} = 27.47$ (see Supplemental Material~\cite{Supplemental} for a derivation of the standard expressions for the $\pi$-pulse average energies) which certifies that our method is energy efficient. A similar situation occurs in Fig.~\ref{classicalsignal}. Here for obtaining the same contrast as in the ideal case, top-hat pulses require at least $\Omega = (2\pi)\times 50$ MHz, larger than the maximum value of $\Omega(t)$ in Fig.\ref{classicalsignal} a). This leads a peak power ratio $\approx 38$  while  $E_{[\pi-\rm pulse ]}^{\rm top-hat}/E_{[\pi-\rm pulse]}^{\rm shaped} = 10.63$.

\section{Conclusions} We have considered quantum sensing experiments at high magnetic
fields that lead to high frequency signals and demonstrated that the finite length of standard
top-hat pulses in dynamical decoupling sequences lead to a rapid decrease of sensitivity with
signal frequency. We present a general solution to this problem which allows to significant reductions in the required
microwave power.  Our method is general and applicable to any magnetic defect.

\section{acknowledgments}
This work was supported by the ERC Synergy grant BioQ (grant no 319130), the EU STREP project HYPERDIAMOND and the DFG CRC 1279. The authors acknowledge support by the state of Baden-W\"urttemberg through bwHPC and the German Research Foundation (DFG) through grant no INST 40/467-1 FUGG. J. C. acknowledges Universit\"at Ulm for a Forschungsbonus.

\begin{appendix}
\renewcommand\thefigure{\thesection.\arabic{figure}}
\setcounter{figure}{0}
\section{NV-nucleus Hamiltonian}\label{appfirst}
The Hamiltonian~(\ref{rfnucleus})  in the rotating frame of the electron-spin free-energy term, $DS_z^2 -\gamma_e B_z S_z$, 
reads
\begin{equation}\label{forsimulations}
H = -\omega_n \ \hat{\omega}_n \cdot \vec{I} + \frac{1}{2}\sigma_z \  \vec{A}\cdot\vec{I} + \frac{\Omega}{2} (|1\rangle \langle 0| e^{i\phi} + |0\rangle \langle 1| e^{-i\phi}).
\end{equation}
Here, we have eliminated any electron spin component containing the $|-1\rangle$ state because, as the MW driving is tuned with the $|0\rangle \leftrightarrow |1\rangle$ transition, the $|-1\rangle$ component gets no populated as well as fast counter rotating terms of the MW driving. In this manner, and as it is commented in the text, we will use this Hamiltonian as the starting point for our numerical simulations. The $\vec{\omega}_n$ vector is  $\vec{\omega}_n = (-\frac{1}{2} A_x, -\frac{1}{2} A_y, \omega_{\rm L} -\frac{1}{2} A_z)$, with $\vec{A} = \frac{\mu_0\gamma_e \gamma_n}{2|\vec{r}|^3}[ \hat{z} - 3 \frac{(\hat{z} \cdot \vec{r}) \vec{r}}{|\vec{r}|^2}]$ being the hyperfine vector (note that we are assuming dipole-dipole interactions between the NV and the nuclear spin) $\omega_n = |\vec{\omega}_n|$ is the resonance frequency of the nucleus which is shifted from the Larmor frequency because of the hyperfine field, and $\hat{\omega}_n = \vec{\omega}_n/\omega_n$.

In a new rotating frame with respect to (w.r.t.), both, $-\omega_n \ \hat{\omega}_n \cdot \vec{I}$ and to the MW driving, one can find that 
\begin{equation}
H = \frac{F_z(t)\sigma_z}{2}  \vec{I}\cdot\bigg[ (\vec{A} -  \vec{A}\cdot \hat{\omega}_n \hat{\omega}_n) \cos{(\omega_n t)} + \hat{\omega}_n\times\vec{A} \sin{(\omega_n t)} + \vec{A}\cdot \hat{\omega}_n \hat{\omega}_n\bigg],
\end{equation}
where $F_z(t)$ is the modulation function, see~\cite{Lang17}, that appears as a consequence of the applied $\pi$-pulses. As $F_z(t)$ will alternate between $+1$ and $-1$ the constant term $\vec{A}\cdot \hat{\omega}_n \hat{\omega}_n$ can be averaged out. Then, the above Hamiltonian can be written as  
\begin{equation}\label{Seffective}
H = \frac{F_z(t)}{2} \ \sigma_z \bigg[ A^\perp_x I_x \cos{(\omega_n t)} + A^\perp_y I_y \sin{(\omega_n t)} \bigg],
\end{equation}
where $A^\perp_x = |\vec{A} -  \vec{A}\cdot \hat{\omega}_n \hat{\omega}_n | = A^\perp_y =|\hat{\omega}_n\times\vec{A}|$, the $\hat{x}$ and $\hat{y}$ directions are  $\hat{x} = (\vec{A} -  \vec{A}\cdot \hat{\omega}_n \hat{\omega}_n)/A^\perp_x$, $\hat{y} = \hat{\omega}_n\times\vec{A}/A^\perp_y$, and $I_x = \vec{I}\cdot \hat{x}$,  $I_y = \vec{I}\cdot \hat{y}$. In this manner, Eq~(\ref{Seffective}) coincides with Hamiltonian~(\ref{basic}). 

\section{The case of a classical field}\label{appsecond}
 If we want to detect classical signals of the form $\vec{B}_{\rm s} \cos{(\omega_{\rm s} t)}$, instead of the initial Hamiltonian in Eq.~(\ref{rfnucleus}), we have to consider
\begin{equation}
H = DS_z^2 - \gamma_e B_z S_z + \Omega_{\rm s} S_z \cos{(\omega_{\rm s} t)} + \sqrt{2} \Omega S_x \cos(\omega t -\phi), 
\end{equation}
where the target signal Rabi frequency is  $\Omega_{\rm s} = \gamma_e \vec{B}_{\rm s}\cdot \hat{z}$. The other components ($\hat{x}$ and $\hat{y}$) of the classical signal field $\vec{B}_{\rm s} \cos{(\omega_{\rm s} t)}$ can be averaged out because $\omega_{\rm s}$ is not on resonance with any of the two possible NV electron spin transitions. More specifically, if we assume that  $\vec{B}_{\rm s} \cos{(\omega_{\rm s} t)}$ is generated by a spin cluster, $\omega_{\rm s}$ would be on the range of several MHz for $B_z$  around a few teslas, while NV transitions require several the Gigahertz to be excited.

In the rotating frame of  $DS_z^2 + |\gamma_e| B_z S_z $ and setting on resonance the driving frequency $\omega$ with, for example, the $|0\rangle \leftrightarrow |1\rangle$ transition, we have 
\begin{equation}\label{SMclassical}
H =  \frac{\Omega_s}{2} \sigma_z \cos{(\omega_{\rm s} t)} + \frac{\Omega}{2} (|1\rangle \langle 0| e^{i\phi} + |0\rangle \langle 1| e^{-i\phi}),
\end{equation}
once we have eliminated constants and any term including the $|-1\rangle$ NV spin component. The use of a standard scheme, as the Hartmann-Hahn double resonance condition~\cite{Hartmann62}, to make interact the NV with the classical field is limited to low values of $\omega_s$. This is because it is required to hold $\omega_s = \Omega$ and the driving power, here reflected in the value of $\Omega$, could be limited. The latter can be easily seen by noting that, in a rotating frame with respect to $ \frac{\Omega}{2} (|1\rangle \langle 0| + |0\rangle \langle 1|)$ where we set $\phi=0$ and $\Omega$ is constant, the above Hamiltonian is 
\begin{equation}
H = \frac{\Omega_s}{4} (|+\rangle\langle -| e^{i\Omega t} + |-\rangle\langle +|e^{-i\Omega t} )(e^{i\omega_s t} + e^{-i\omega_s t} ).
\end{equation}
where $|\pm\rangle = \frac{1}{\sqrt{2}}(|\uparrow\rangle \pm i|\downarrow\rangle)$. Now one can see that, unless the condition $\Omega = \omega_s$ holds, the previous Hamiltonian is entirely time-dependent and would average to zero because of the rotating wave approximation.

Hence, one should consider the use of pulsed schemes. In this case, Eq.~(\ref{SMclassical}) can be written in the rotating frame of the MW driving $\frac{\Omega}{2} (|1\rangle \langle 0| e^{i\phi} + |0\rangle \langle 1| e^{-i\phi})$ with $\Omega$ applied stroboscopically, as
\begin{equation}
H =  \frac{\Omega_s}{2} F_z(t) \sigma_z \cos{(\omega_s t)}.
\end{equation}
Now, as $ F_z(t) = \sum_l f_l \cos{(l \omega_{\rm m} t)}$, and in the case of $k \omega_{\rm m} = \omega_s$ (i.e. resonance condition for the $k$th harmonic) the above Hamiltonian after eliminating fast rotating terms is 
\begin{equation}
H = \frac{\Omega_s}{4}f_k \ \sigma_z,
\end{equation}
which is Eq.~(\ref{forclassical}).

\section{Integrating $F_z(u)$}\label{appthird}
The condition in Eq.~(\ref{condition}) when applied to the solution in Eq.~(\ref{ansatz}) leads to the two following integrals 
\begin{eqnarray}
&&\int_{t_1}^{t_2} \cos{\bigg[\frac{\pi}{t_\pi} (u - t_1)\bigg]} \ \cos{\bigg[k\frac{2\pi}{T} u\bigg]} \ du \nonumber\\
&=& t_{\pi} \bigg[ \frac{1}{(2 \alpha +1)\pi} + \frac{1}{(2 \alpha -1)\pi}\bigg],
\end{eqnarray}
and
\begin{eqnarray}
&&\int_{t_1}^{t_2} \exp{\bigg[-\frac{(u-t_p)^2}{2c^2}\bigg]} \sin{\bigg[\frac{2\pi \alpha }{t_\pi}(u - t_1)\bigg]}\ \cos{\bigg[k\frac{2\pi}{T} u\bigg]} \ du \nonumber \\&\approx& c\sqrt{\frac{\pi}{2}}\bigg[1 - \exp{\bigg(\frac{-8\alpha^2\pi^2}{\gamma^2}\bigg)}\bigg]
\end{eqnarray}
with $c=\frac{t_\pi}{\gamma}$. Note that for solving the last integral we have used the following relations
\begin{eqnarray}
&\int_{-\infty}^{+\infty} e^{-(x-b)^2/2c^2} dx = \sqrt{2} |c| \sqrt{\pi} \nonumber\\
&\int_{-\infty}^{+\infty} e^{-ax^2} \cos{(bx)} dx = \sqrt{\frac{\pi}{a}} e^{-b^2/4a}.
\end{eqnarray}

\end{appendix}

\end{document}